# Laboratory Studies on the Carbon Kinetic Isotope Effects on the Production Mechanism of Particulate Phenolic Compounds Formed by Toluene Photooxidation: A Tool to Constrain Reaction Pathways

Satoshi Irei,*[1a] Jochen Rudolph,[1] Lin Huang,[2] Janeen Auld,[1b] Fabrice Collin,[1c] and Donald Hastie[1]

[1]Centre for Atmospheric Chemistry and Department of Chemistry, York University, 4700 Keele St., Toronto, Ontario, M3J 1P3, Canada

[2]Climate Research Division, Atmospheric Science and Technology Directorate, Science and Technology Branch, Environment Canada, 4905 Dufferin St., Toronto, Ontario, M3H 5T4, Canada

[a]Present address: Department of Chemistry, Biology, and Marine Science, Faculty of Science, University of the Ryukyus, 1 Senbaru, Nishihara, Okinawa 903-0213, Japan

[b]Present address: Department of Chemistry and Biochemistry, University of Windsor, Windsor, ON, N9B 3P4, Canada

[c]Present address: PHARMA-DEV, UMR 152 IRD-UPS, Université Paul Sabatier, 118 route de Narbonne, 31400 Toulouse, France.

*Corresponding author: University of the Ryukyus, Japan. Tel: +81-98-850-2696; fax: +81-98-850-2579; e-mail: satoshi.irei@gmail.com.

**ABSTRACT:** In this study, we examined compound-specific stable carbon isotope ratios for phenolic compounds in secondary organic aerosols (SOA) formed by photooxidation of isotope-label free toluene. SOA generated by photooxidation of toluene using a continuous-flow reactor and an 8-m$^3$ indoor smog chamber was collected on filters, which were extracted with acetonitrile for compound-specific analysis. Eight phenolic compounds were identified in the extracts using a gas chromatograph coupled with a mass spectrometer, and their compound-specific stable carbon isotope ratios were determined using a gas chromatograph coupled with a combustion furnace followed by an isotope ratio mass spectrometer. The majority of products, including methylnitrophenols and methylnitrocatechols, were isotopically depleted by 5‰–6‰ compared to the initial isotope ratio of toluene, whereas the isotope ratio for 4-nitrophenol remained identical to that of toluene. Based on the reaction mechanisms proposed in previous reports, stable carbon isotope ratios of these products were calculated. By comparing the observed isotope ratios with the predicted isotope ratios, we explored possible production pathways for the particulate phenolic compounds.

1. INTRODUCTION

Airborne particulate organic matter or organic aerosol (OA), which is a predominant component of atmospheric aerosols,[1] has been of great interest in the past two decades owing to its possible association with adverse health effects[2] and with the global budget of radiative forcing.[3] However, these associations are not clearly understood. One issue is the limited quantitative information available on secondary OA (SOA), which is formed by oxidative conversion of volatile organic compounds in the atmosphere. These difficulties are attributed to an insufficient understanding of chemical reaction mechanisms leading to the formation of SOA. A newly developed approach that can provide novel insights into SOA formation is studying the isotope fractionation of isotope label-free compounds using high precision isotope ratio measurements.

Highly precise stable isotope ratio measurements of compounds containing stable isotopes with a natural abundance can be used to increase our understanding of complex mixing and reaction mechanisms. For example, Rudolph[4] has demonstrated that isotope fractionations that occur during reactions of volatile organic compounds (VOCs) with oxidants can provide valuable information on the extent of chemical reactions for VOCs in the atmosphere, which, owing to atmospheric dilution with background air, is difficult to obtain based on concentration measurements alone. Besides this work, several reports have showed that stable isotope ratio measurement provides useful indications regarding atmospheric photochemistry.[5–9] One advantage of this technique is that the measurements do not require labeled reactants with stable isotopes. Instead, relatively small changes in natural isotope abundance are used to evaluate the extent of reaction processing. This allows isotope effects determined from laboratory studies to be directly applied to ambient observations. A kinetic isotope effect (KIE), which is an isotope fractionation during a chemical reaction, is reaction dependent; thus, measuring changes in isotope ratios during chemical reactions will provide additional insights into reaction mechanisms, such as SOA formation. To our knowledge, few studies have examined isotope fractionation during SOA formation.[10–12] Our previous studies focused on stable carbon isotope ratios ($\delta^{13}$C) for total SOA carbon produced by the photooxidation of toluene in a flow reactor and a smog chamber under a variety of experimental conditions.[10,11] In these studies, we found that the KIE for the initial reaction step explained the $\delta^{13}$C variation of SOA within the uncertainties of the measurements. In this report, our objective was to increase our understanding of the $\delta^{13}$C balance of SOAs using a newly developed compound-specific stable carbon isotope analysis for nitrophenols.[13]

Nitro and nitrohydroxy aromatic compounds have been detected in gas-phase products[14,15] as well as SOA[16–19] produced by reactions of aromatic hydrocarbons with OH radicals in laboratory studies. Based on these laboratory studies, production mechanisms have been postulated. Thus, the application of compound-specific label-free isotope measurement to these compounds can be used to constrain reaction channels in reaction mechanism studies. In this report, the results are discussed based on postulated processes for the formation of phenolic products with the estimated carbon KIEs for the individual reaction steps.

## 2. EXPERIMENTAL

**2.1. Flow reactor and smog chamber experiments.** Laboratory studies of SOA formed by the photooxidation of toluene were conducted with a continuous-flow reactor and a smog chamber. The former has features that production of SOA is theoretically unlimited, but extent of toluene reaction is limited due to the low residence time. Meanwhile, the latter apparatus has features that extent of toluene reaction goes on as long as UV lights are on (i.e, capability of high toluene turnover), but production of SOA is limited due to the limited volume of the apparatus. SOA samples formed under a wide range of toluene turnover conditions were collected. Details on the flow reactor and smog chamber experiments were described previously.[10,11,19,20] Briefly, a mixture of synthetic air, NO (BOC gases), toluene (purity > 99%, Sigma-Aldrich Canada), and synthesized isopropyl nitrite was irradiated with UV light to generate OH radicals. The initial mixing ratios of NO, toluene, and isopropyl nitrite were 5, 40, and ppmv for the flow reactor experiments and 3.5, 0.5, and 0.7 ppmv for the smog chamber experiments, respectively (Table 1). The initial $\delta^{13}C$ values of toluene determined based on conventional off-line isotope ratio measurements were –26.72‰ ± 0.01‰ and –27.05‰ ± 0.02‰ for the flow reactor experiments and the smog chamber experiments, respectively. Although toluene used for these experiments was from the same batch, the slightly heavier $\delta^{13}C$ for the flow reactor experiments (i.e., -26.72‰) was due to the small isotope fractionation during partial evaporation from bulk toluene. A detailed description of the evaporative isotope fractionation is discussed previously.[10,21] This isotope fractionation is only one twentieth of the isotope fractionation observed for the toluene reaction with OH radicals, and did not significantly impact our observations. No seed particles were introduced in the flow reactor experiments, while dried ammonium sulfate particles were introduced as SOA seeds in the smog chamber experiments ($2 \times 10^5$ particles cm$^{-3}$ with a monomodal distribution and a maximum diameter of 100 nm). The gas mixture was irradiated using a high-pressure xenon arc lamp (1000 W, model C-50, Oriel Optical Corp.) for the flow reactor experiments. For the smog chamber experiments, the gas/seed particle mixture was irradiated by 24

UV lights (40 W, 350–400-nm wavelength, model F40BL, Philips). The nucleation of SOA was confirmed using a condensation nuclei counter (Model 3020; TSI Inc., Shoreview, MN) during the flow reactor experiments. During the smog chamber experiments, the growth of seed particles (i.e., SOA production) was confirmed based on size distribution measurements using a combination of a condensation nuclei counter and a differential mobility analyzer (TSI-3760 and TSI-3071; TSI Inc., Shoreview, MN). The extent of the toluene reaction was determined by measuring its concentration using a gas chromatograph coupled with a flame ionization detector. In the flow reactor and smog chamber experiments, the SOA was collected on 25-mm-o.d. and 47-mm-o.d. polytetrafluoroethylene (PTFE)-coated glass fiber disc filters (Pallflex Fiberfilm; Pall Corp., Port Washington, NY), respectively. The filters were used without pre-cleaning. Filter sampling was summarized in Table 1. Sampling flow rates for the flow reactor and the smog chamber experiments were 6.3 and 12 L min$^{-1}$, respectively. These corresponded to linear velocities of 1700 and 1300 cm min$^{-1}$, respectively. These linear velocities were smaller than that used for usual high-volume sampling (~2000 cm min$^{-1}$), therefore, the blow-off effect of collected SOA was lower than that for typical ambient particulate filter sampling using a high-volume air sampler.

Table 1. Summary of initial experimental conditions and SOA filter sampling

|  | *Flow Reactor Experiments* <br> *FR 1 - FR 5* | *Smog Chamber Experiments* <br> *SC I - SC III* |
|---|---|---|
| ***Initial experimental conditions*** | | |
| NO (ppmv) | 5 | 2.5 |
| Toluene (ppmv) | 20 - 40 | 0.5 |
| IPN (ppmv) | 230 | 7 |
| Seed particles (particles cm$^{-3}$) | 0 | 3000 |
| ***Filter sampling*** | | |
| Number of filter sample collected | 5 | 3 |
| Sampling duration (h) | 2.4-19.0 | 5.03 - 5.22 |
| Sampling flow rate (L min$^{-1}$) | 6.3 | 12 |
| Sampled volume (m$^{-3}$) | 0.1 - 0.9 | 3.6 - 3.8 |

**2.2. Analysis of filter samples.** Briefly, a total of 5 or 24 µg C of catechol, 9 or 26 µg C of 2-nitrophenol, 6 or 23 µg C of 4-ethylresorcinol, and 52 or 16 µg C of 2,6-dimethyl-4-nitrophenol (> 99% purity; Sigma-Aldrich, Oakville, ON, Canada) were spiked to a one-eighth segment of a filter sample as internal standards (IS) for quantitative analysis of phenolic compounds and routine isotope measurements. Hereafter, these internal standards are referred to as IS1, IS2, IS3, and IS4, respectively. The filter was then extracted with 2–3 mL of HPLC-grade acetonitrile (Sigma-Aldrich, Oakville, ON, Canada) three times by sonication for 3–5 min in a 60-mL glass jar (wide-mouth amber glass jars; Chromatographic Specialties Inc., Brockville, ON, Canada) pre-silanized with dimethyldichlorosilane (Sigma-Aldrich, Oakville, ON, Canada). The extracts were filtered using a PTFE syringe filter (25-mm o.d., 0.45-μm pore disk filter; Chromatographic Specialties Inc.) and then combined. The combined extracts were then concentrated to ~0.1 mL under a gentle flow of analytical-grade nitrogen (Praxair Canada Inc., Mississauga, ON, Canada) in 5-mL vials (Reacti-Vial; Pierce Chemical Co., Rockford, IL). After the volume reduction, 1–8 μg C of $C_{17}$ to $C_{19}$ n-alkanes dissolved in acetonitrile was added to determine the final volume of the concentrated extract. Prior to the measurement, an aliquot of the concentrated extract was derivatized with N,O-bis(trimethylsilyl)trifluoroacetamide (99% purity, BSTFA; Regis Technologies, Morton Grove, IL). The derivatized sample was analyzed using a gas chromatograph coupled with an ion-trap mass spectrometer, GC-MS (CP 3800 and Saturn 2000; Varian, Palo Alto, CA), for compound identification. The derivatized phenolic compounds were analyzed using a gas chromatograph (6890A, Agilent Technologies, Santa Clara, CA) coupled with a combustion furnace followed by an isotope ratio mass spectrometer (IsoPrime, GV Instruments, Manchester, UK) for isotope ratio and concentration measurements. The majority of extracts were analyzed on the day of preparation; otherwise, the extracts were stored in a refrigerator. All extracts were analyzed within one week of extraction. This methodology was thoroughly evaluated using 11 reference phenolic compounds [12] and recovery yields of these compounds were better than 80%. Their retention times in the GCC–IRMS chromatograms and reference $\delta^{13}C$ values obtained using off-line combustion and dual-inlet isotope ratio analysis were listed in Table S-1.

The BSTFA derivatization (referred to above) replaces an active hydrogen atom of the OH group with a trimethylsilyl (TMS) group, causing a defined isotopic shift from the original $\delta^{13}C$ value. This isotopic shift was corrected as follows to obtain the $\delta^{13}C$ values of the TMS-free compounds ($\delta^{13}C_{free}$):

$$\delta^{13}C_{free} = \frac{\#C_{deriv}}{\#C_{free}} \times \delta^{13}C_{deriv} - \frac{\#C_{TMS}}{\#C_{free}} \times \delta^{13}C_{TMS} \qquad (1)$$

where $\delta^{13}C_{deriv}$, $\delta^{13}C_{TMS}$, $\#C_{deriv}$, $\#C_{TMS}$, and $\#C_{free}$ are the $\delta^{13}C$ values for the derivatized compound, the $\delta^{13}C$ value for the TMS group, the total number of carbon atoms in the derivatized molecule, the number of carbon atoms in the TMS group, and the number of carbon atoms in the TMS-free compound. The carbon atom numbers were determined according to the stoichiometry of the derivatization reaction as long as the original compound was identified. We used –49.94‰ ± 0.33‰ (±standard error) for $\delta^{13}C_{TMS}$ because the BSTFA used here was the same batch as reported previously.[13]

## 3. RESULTS AND DISCUSSION
**3.1. Product Identification.** GCC–IRMS analysis of the PTFE blank filter extracts showed some contaminants in the extracts. Although the retention times of the contaminants differed from those of the target compounds, they differed from the retention times of the internal standards IS3 and IS4 by only ~0.4 min in some cases. The contaminant levels varied from sample to sample, and the measurement results for these internal standards may be influenced to some extent by these contaminant peaks.

GCC–IRMS analysis revealed significant abundances of eight product peaks in the flow reactor samples (Figure 1), while only one prominent product peak was observed in one of the smog chamber samples. We identified seven products based on GC–MS spectral patterns, retention times and the abundance of the *m/z* 46 peak in the GCC–IRMS spectrum. The *m/z* 46 peak is an indication that the product contains a nitro group. Table 2 summarizes the identified products and the methods used for identification. The following is a brief description of the product identification.

Retention times and mass spectra for peaks 2 and 3 agreed with those for 3-methyl-4-nitrophenol and 2-methyl-4-nitrophenol, respectively. Therefore, these peaks represented 3-methyl-4-nitrophenol and 2-methyl-4-nitrophenol, respectively. The most abundant peak in the flow reactor extracts was 2-methyl-4-nitrophenol, which was the only product identified in the smog chamber extracts. The mass spectrum for peak 1 showed remarkably high signals at *m/z* 73 and 45, indicative of a TMS derivative of a substance containing OH and/or COOH group(s). High signal intensities at *m/z* 135, 150, and 196 were similarly observed for the reference, 2-nitrophenol. Additionally, a high abundance in the *m/z* 46 trace in the GCC–IRMS spectrum was observed for peak 1, suggesting that the product contained at least one nitro group. Based on this, peak 1 was most likely a structural isomer of nitrophenol. Analysis of 3- and 2-nitrophenols revealed that their retention times did not match the retention time for peak 1. Therefore, we identified peak 1 as 4-nitrophenol.

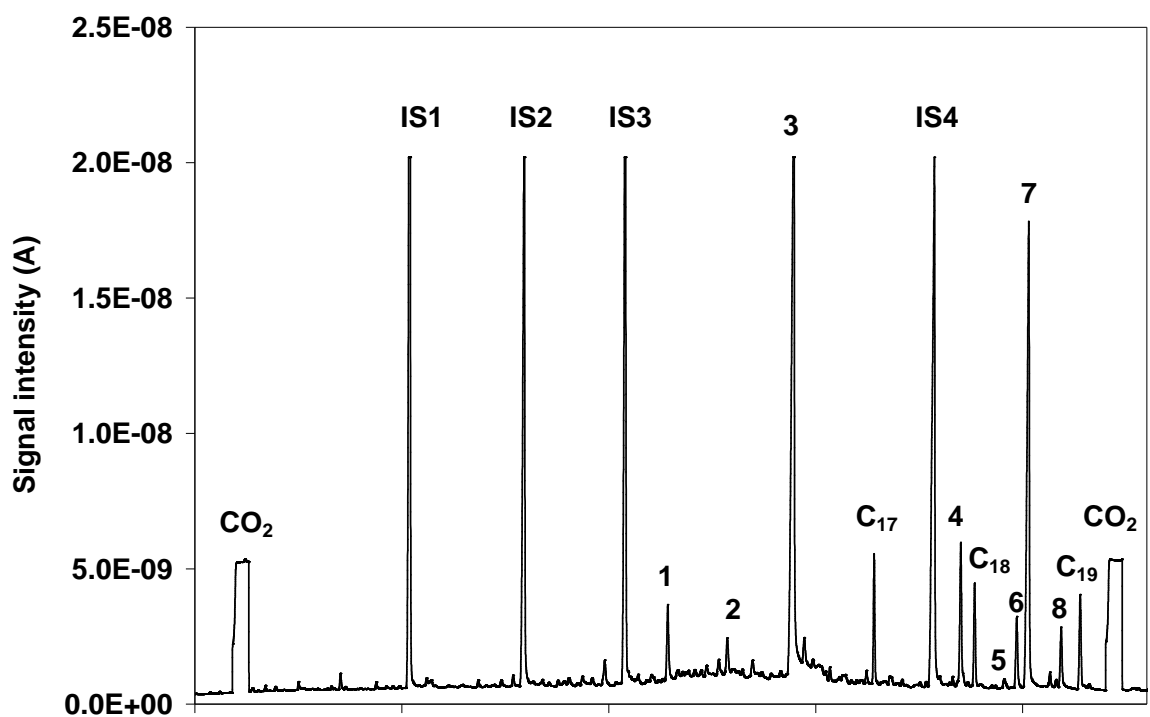

**Figure 1.** Chromatogram of the $m/z$ 44 trace obtained based on GCC–IRMS analysis of a SOA extract. IS1, IS2, IS3, IS4, $C_{17}$, $C_{18}$, and $C_{19}$ are catechol, 2-nitrophenol, 4-ethylresorcinol, 2,6-dimethyl-4-nitrophenol, heptadecane, octadecane, and nonadecane, used as internal or volumetric standards. The numbered peaks are products found in the SOA extracts. See Table 2 for product identification.

**Table 2. Identified products and methods of their identification**

| Peak no. | Retention time (min) | Identified product (abbreviation) | Method of identification[a] |
|---|---|---|---|
| 1 | 48.9 | 4-nitrophenol (4NP) | MS and $m/z$ 46 |
| 2 | 51.4 | 3-methyl-4-nitrophenol (3Me4) | RT, MS, and $m/z$ 46 |
| 3 | 54.0 | 2-methyl-4-nitrophenol | RT, MS, and $m/z$ 46 |
| 4 | 60.7 | (2Me4)<br>methylnitrocatechol isomer 1<br>(MeNC1) | MS and *m/z* 46 |
| 5 | 62.5 | unknown | Unsuccessful |
| 6 | 63.0 | methylnitrocatechol isomer 2<br>(MeNC2) | MS and *m/z* 46 |
| 7 | 63.4 | methylnitrocatechol isomer 3<br>(MeNC3) | MS and *m/z* 46 |
| 8 | 64.8 | nitrocatechol isomer<br>(NC) | MS and *m/z* 46 |

*a*RT, MS, and *m/z* 46 denote the identification method of retention time matching, mass spectral pattern, and detection of the high signal peak in the *m/z* 46 trace chromatogram (identification of the nitro group) based on GCC-IRMS analysis, respectively.

Peaks 4, 6, 7, and 8 showed very similar mass spectra, suggesting that they are isomeric structures. However, the retention times of these peaks did not correspond to reference compounds. Peaks 4, 6, and 7 had the largest mass at *m/z* 313, and peak 8 had the largest mass at *m/z* 298. These ions were likely molecular ions. According to the analysis of reference compounds, the signals at *m/z* 73 and 45 are indicators of TMS derivatives. For peaks 4, 6, and 7, the relative abundances of these *m/z* signals were larger than those for the reference nitrophenols but similar to those for the reference catechols, indicative of dihydroxy compounds. Furthermore, remarkably high signals in the chromatogram of the *m/z* 46 trace were indicative of a nitro group. Based on these results, we identified peaks 4, 6, and 7 as methylnitrocatechol isomers (MM = 313) and peak 8 as a nitrocatechol isomer (MM = 298). Unfortunately, peak 5 could not be identified.

**3.2. Yields and $\delta^{13}$C values of specific products in SOA.** $\delta^{13}C_{free}$ values and yields of identified products in the flow reactor samples for various extents of the toluene reaction are listed in Table 3. All $\delta^{13}C_{free}$ values were calculated according to Eq. (1) with the reference $\delta^{13}C_{TMS}$ value ± standard error of –49.94‰ ± 0.33‰ determined for the TMS group of BSTFA in this batch.[19] Yield is defined as carbon mass of the identified product relative to the carbon mass of toluene reacted. Carbon masses were determined according to the internal calibration for structurally similar reference compounds listed in Table S1. The yield measurements of the flow reactor samples showed that 2-methyl-4-nitrophenol accounted for 0.3%–3% of reacted toluene, followed by methylnitrocatechol isomer 3 (0.1%–0.6%), 4-nitrophenol and methylnitrocatechol isomer 1 (0.1%–0.2%), and other products (<0.1%). These yields increased as the extent of toluene reaction progressed. Based on the dependence between total SOA carbon yields and the extent of toluene reaction processing,[10] carbon mass fractions of 4-nitrophenol, 3-methyl-4-nitrophenol, 2-methyl-4-nitrophenol, methylnitrocatechol isomer 1, isomer 2, isomer 3, and nitrocatechol in SOA were 3.1% ± 0.5%, 3.2% (no error available), 54% ± 19%, 1.2% ± 0.2%, 0.9% (no error available), 7.6% ± 1%, and 0.5% ± 0.1%, respectively. The $\delta^{13}C_{free}$ values for all products, with the exception of 4-nitrophenol, in the flow reactor samples were between –29‰ and –37‰, indicating that these products were significantly depleted in $^{13}C$ relative to the starting material, toluene ($\delta^{13}C$ for toluene ≈ –27‰). In contrast, the $\delta^{13}C_{free}$ values for 4-nitrophenol ranged from –25‰ to –28‰, indicating that $^{13}C$ in 4-nitrophenol was slightly enriched or similar to that in initial toluene. Comparison of photolysis rates of nitrophenol ($3\times10^{-4}$ to $7\times10^{-4}$ s$^{-1}$)[22] with the product of the rate constants for photooxidation of nitrophenols ($\sim5\times10^{-12}$ cm$^3$ molecule$^{-1}$ s$^{-1}$)[22] multiplied by estimated OH concentration in the flow reactor based on the toluene loss ($1\times10^9$ to $1\times10^{10}$ molecules cm$^{-3}$) and the residence time (0.6 to 3.2 min)[10] suggests that the photolysis accounted for less than 10%. Therefore, the wide range of observed $\delta^{13}C$ suggests that the products and their precursors underwent chemical reactions with different carbon KIEs.

**Table 3. $\delta^{13}C_{free}$ values and carbon yields[a] (± standard deviation from triplicate measurements) of specific products identified in the SOA from the flow reactor experiment (FR 1 – FR 5)**

| FR 5<br>9 ± 1% toluene reaction | Carbon yield (%) | N/A[b] | N/A[b] | 0.256 ± 0.005 | 0.015 ± 0.000 | N/A[b] | 0.064 ± 0.001 | 0.024 ± 0.001 |
|---|---|---|---|---|---|---|---|---|

| | FR 4 | |
|---|---|---|
| | 10 ± 1% toluene reaction | |
| $\delta^{13}C$ (‰) | Carbon yield (%) | $\delta^{13}C$ (‰) |
| -26.4 ± 0.3 | 0.019 ± 0.001 | N/A[b] |
| N/A[b] | N/A[b] | N/A[b] |
| -33.1 ± 0.7 | 0.30 ± 0.01 | -29.8 ± 0.4 |
| -32.85 ± 0.01 | 0.015 ± 0.000 | -32.4 ± 1.0 |
| N/A[b] | N/A[b] | N/A[b] |
| -32.7 ± 0.5 | 0.061 ± 0.001 | -33.9 ± 0.3 |
| -33.1 ± 1.2 | 0.022 ± 0.001 | -32.4 ± 0.6 |

| FR 2 | | FR 3 | |
| --- | --- | --- | --- |
| 27 ± 2% toluene reaction | | 21 ± 2% toluene reaction | |
| $\delta^{13}C$ (‰) | Carbon yield (%) | $\delta^{13}C$ (‰) | Carbon yield (%) |
| -28 ± 1 | 0.170 ± 0.004 | -25 ± 2 | 0.133 ± 0.008 |
| -34.0 ± 0.2 | 0.083 ± 0.002 | -32 ± 3 | 0.077 ± 0.006 |
| -30.5 ± 0.3 | 2.69 ± 0.02 | N/A[b] | 3.20[c] |
| -31.6 ± 0.6 | 0.110 ± 0.003 | -30.9 ± 0.6 | 0.065 ± 0.002 |
| -32.4 ± 0.6 | 0.058 ± 0.002 | -35.4 ± 0.9 | 0.041 ± 0.001 |
| -33.98d | 0.48 ± 0.01 | -32.9 ± 0.3 | 0.28 ± 0.01 |
| -30.7 ± 0.6 | 0.051 ± 0.001 | -32.1 ± 0.6 | 0.034 ± 0.001 |

| Product identified | FR 1 17 ± 2% toluene reaction | |
|---|---|---|
| | δ$^{13}$C (‰) | Carbon yield (%) |
| 4ntrphen | -26.7 ± 0.7 | 0.148 ± 0.004 |
| 3Me4 | -36.5 ± 1.8 | 0.039 ± 0.001 |
| 2Me4 | -32.6 ± 0.3 | 3.23 ± 0.02 |
| Mentrctchl 1 | -32.2 ± 0.3 | 0.159 ± 0.005 |
| Mentrctchl 2 | -33.7 ± 0.6 | 0.079 ± 0.002 |
| Mentrctchl 3 | -32.5 ± 0.5 | 0.64 ± 0.01 |
| ntrctchl | -30.4 ± 1.0 | 0.083 ± 0.003 |

[a]Carbon yield is defined as carbon mass of identified product divided by reacted carbon mass of toluene. [b]No data available. [c]No standard deviation is available due to only one available measurement.

Due to the very low nitrophenol concentrations, the analysis of smog chamber samples resulted in only one successful detection of 2-methyl-4-nitrophenol from the SC II experiment. Determined yield and δ$^{13}$C$_{free}$ values were 0.042% and –29.3‰, respectively. Compared to the total SOA carbon yields for the smog chamber experiments, 2-methyl-4-nitrophenol accounted for only ~0.2% of SOA carbon. This is indicative of the different chemical compositions of SOA formed in the flow reactor experiments compared to that in the smog chamber experiments.

The δ$^{13}$C$_{free}$ values for the identified products were compared with the estimated δ$^{13}$C value for total SOA carbon, which is equivalent to the δ$^{13}$C value for the sum of all products.[9,10] The comparison is discussed below in terms of the difference between the δ$^{13}$C$_{free}$ value for the product and the predicted δ$^{13}$C value for the SOA carbon at the same extent of toluene reaction (i.e., δ$^{13}$C$_{free}$ minus predicted δ$^{13}$C for the SOA carbon), denoted as Δδ$^{13}$C hereafter (Table 4). The Δδ$^{13}$C values for all products excluding 4-nitrophenol were randomly scattered around zero, and the values were comparable in magnitude with standard errors based on triplicate measurements. The average $\Delta\delta^{13}C$ for each compound (excluding 4-nitrophenol and methylnitrocatechol isomer 2) over the five flow reactor experiments was not significantly different from zero at the 95% confidence level. Thus, we concluded that for all products identified here (excluding 4-nitrophenol and methylnitrocatechol isomer 2), there was no significant difference between the $\delta^{13}C$ value predicted by a mass balance and the observed $\delta^{13}C$ value. In contrast, the $\Delta\delta^{13}C$ values for 4-nitrophenol were +6‰ with a standard deviation of 1‰, clearly showing a significant difference.

**Table 4. Difference[a] (± standard error) between the measured $\delta^{13}C_{free}$ value for a specific product and predicted $\delta^{13}C$ value for the sum of all products ($\Delta\delta^{13}C$)**

| Experiment | FR1 | FR2 | FR3 | FR4 | FR5 |
|---|---|---|---|---|---|
| Fraction of toluene reacted | 0.17 | 0.27 | 0.21 | 0.1 | 0.09 |
| | | | (‰) | | |
| 4ntrphen | 5.27±0.49 | 3.84±0.75 | 6.75±0.92 | 5.75±0.36 | n.a.[b] |
| 3Me4 | -4.53±1.10 | -2.36±0.32 | -0.36±1.60 | n.a.[b] | n.a.[b] |
| 2Me4 | -0.62±0.33 | 1.33±0.34 | n.a.[b] | -0.89±0.50 | 2.40±0.37 |
| Mentrctchl1 | -0.20±0.34 | 0.02±0.44 | 0.93±0.44 | -0.67±0.30 | -0.22±0.63 |
| Mentrctchl2 | -1.77±0.48 | -0.78±0.44 | -3.53±0.58 | n.a.[b] | n.a.[b] |
| Mentrctchl3 | -0.57±0.42 | -2.35 | -1.05±0.34 | -0.55±0.42 | -1.72±0.34 |
| ntrctchl | 1.56±0.66 | 0.91±0.46 | -0.29±0.47 | -0.93±0.73 | -0.23±0.46 |

[a]$\Delta\delta^{13}C = \delta^{13}C_{free} - \delta^{13}C$ for the Sum of All Products. [b]Not available.

### 3.3. Comparison with isotope ratios derived from postulated reaction mechanisms

The majority of identified phenolic products have carbon isotope ratios that agree with mass balance-based predictions of the isotope ratios of the sum of all products, indicating the first reaction step is the rate-determining step for production of particulate phenolic compounds. This is somewhat surprising since previously postulated reaction mechanisms for photooxidation of toluene (Figure 2) suggest that the formation of nitrophenols and catechols proceeds via several reaction steps. Therefore, we examined the isotope fractionation expected for postulated reaction channels.

The following is a brief summary of postulated mechanisms for reactions of parent toluene and some of its first-, second-, and third-generation products with OH radicals, $O_2$, and $NO_x$. The reaction of toluene with OH radical initially proceeds through two channels: addition of OH to the aromatic ring (1a channel), accounting for approximately 88% of the initial reaction, and H abstraction from the methyl group (1b channel), accounting for approximately 12% of the initial reaction.[23] The sum of the second-order rate constants for the 1a and the 1b channels is $5.96 \times 10^{-12}$ $cm^3$ $molecule^{-1}$ $s^{-1}$.[24] The 1a channel is known to form a methylhydroxycyclohexadienyl radical, which rapidly decomposes back to the reactants (life time of 0.3 s at 298 K and atmospheric pressure), but also undergoes several forward reactions (2a, 2b, and 2c channels). The 2a channel is believed to yield benzoquinone and ring-opening products.[23] The 2b channel yields cresols (the first-generation product) with yields of 12%–40% of reacted toluene.[17, 25–28] Forstner et al. reported that cresols can act as intermediates for the formation of methylcatechols and methylnitrophenols via the 3a and 3b channels, respectively.[16]

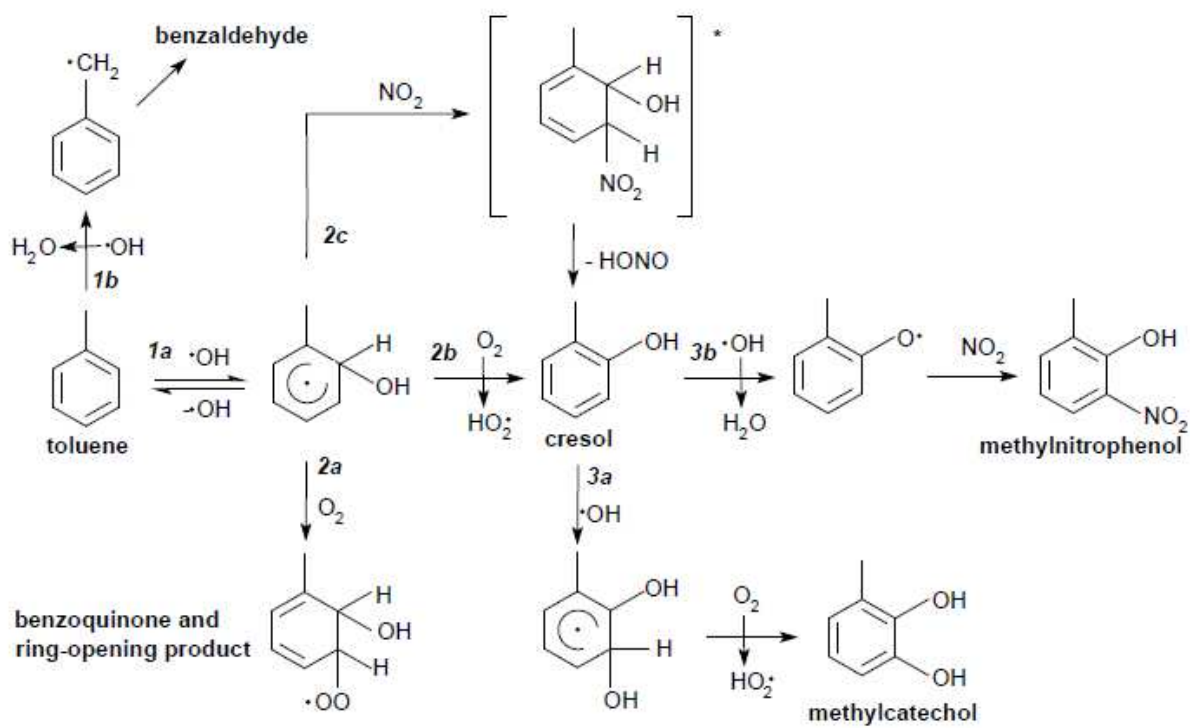

**Figure 2.** Postulated reaction mechanisms for the photooxidation of toluene and the following oxidation reactions[16-19]

The sum of the second-order rate constants for the 2a and 2b channels (the reactions of methylhydroxycyclohexadienyl radical with $O_2$) was approximately $(5.4 \pm 0.6) \times 10^{-16}$ $cm^3$ $molecule^{-1}$ $s^{-1}$ at room temperature.[24] To our knowledge, the branching ratio reported for these channels has not been reported. The methylhydroxycyclohexadienyl radical reacts with $NO_2$ directly at high $NO_2$ concentrations (i.e., > ~400 ppbv). This reaction is depicted by the 2c channel, first postulated by Zellner et al.[29], and produces a methylhydroxynitrocyclohexadienyl radical. The second-order rate constant for the 2c channel is $\sim 4 \times 10^{-11}$ $cm^3$ $molecule^{-1}$ $s^{-1}$.[30] Under our experimental condition with an $O_2$ content of 20%, the 2c channel accounted for 30% and 80% of the loss of methylhydroxycyclohexadienyl radical at $NO_2$ mixing ratios of 1 ppmv (the condition in the smog chamber experiments) and 10 ppmv (the condition in the flow reactor experiments), respectively. The methylhydroxynitrocyclohexadienyl radical is believed to decompose and produce cresol and HONO.[27] Compared to OH addition, H abstraction from the OH substituent of cresol (3b channel) and the methyl substituent is a minor path (~7%) owing to the lower probability of OH attack on the OH substituent of cresol than OH attack to the aromatic ring, but still more favored than abstraction of an H atom from the methyl group.[13] This was experimentally confirmed based on observations by Olariu et al.[31] where hydroxybenzaldehyde, an expected product via H abstraction from the methyl substituent, was not formed. They proposed that the branching ratio of OH addition and H abstraction are 90% and 10%, respectively. The 3a and 3b channels yield methylcatechols and methylnitrophenols, respectively,[31] which may further react with the OH radical. Rate constants for reactions of methylnitrophenols with the OH radical in the gas-phase (4a and 4b channels in Figure 3) range from $3 \times 10^{-12}$ cm$^3$ molecule$^{-1}$ s$^{-1}$ to $7 \times 10^{-12}$ cm$^3$ molecule$^{-1}$ s$^{-1}$.[23] Rate constants for reactions of methylcatechols with the OH radical (5a and 5b channels in Figure 3) range from $1 \times 10^{-10}$ cm$^3$ molecule$^{-1}$ s$^{-1}$ to $2 \times 10^{-10}$ cm$^3$ molecule$^{-1}$ s$^{-1}$.[32] The former rate constants are more than one order of magnitude larger than that for the reaction of toluene with the OH radical, whereas the rate constants for reactions of methylnitrophenols with the OH radical are of similar magnitudes.

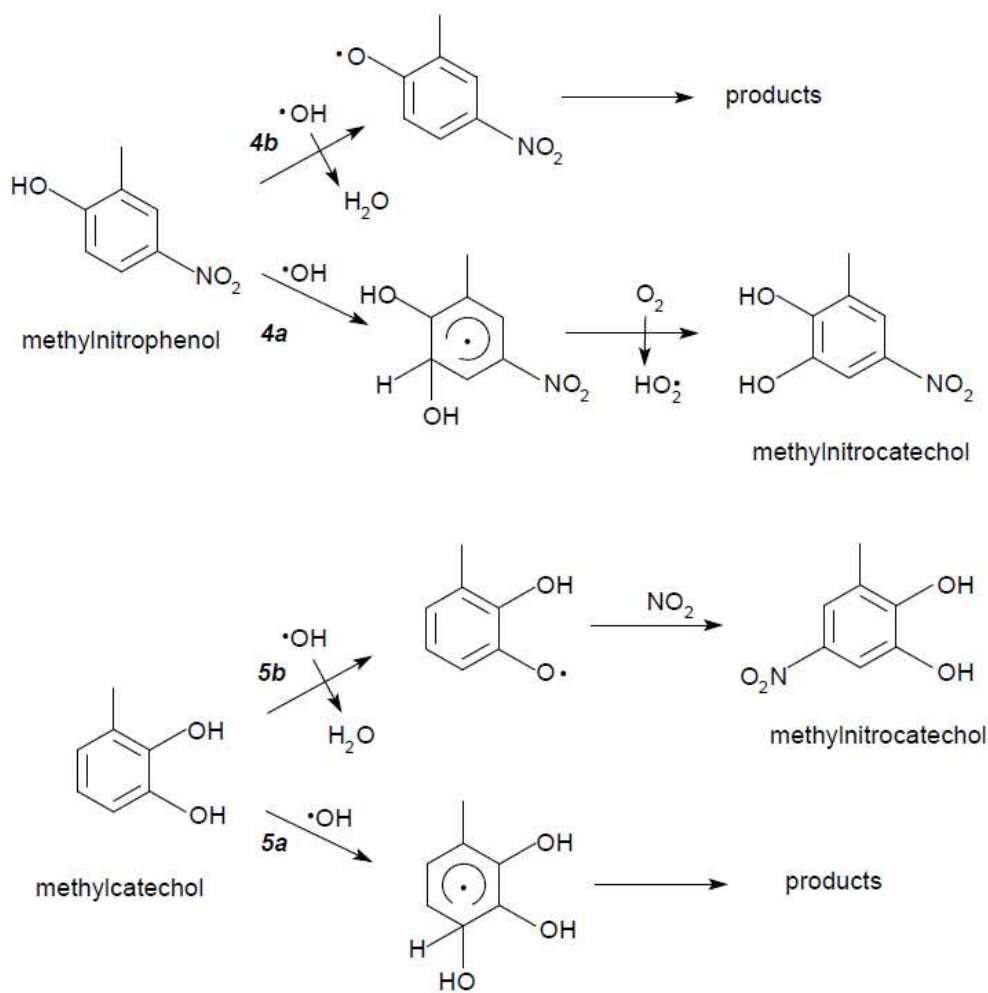

**Figure 3.** Production reaction mechanisms used to calculate $\delta^{13}C$ of methylnitrocatechol

Based on these reaction mechanisms, changes in $\delta^{13}C$ for toluene, methylnitrophenol, and methylnitrocatechol were calculated as a function of the extent of the reaction of toluene with OH radical using the aforementioned rate constants and the expected KIE values, which are determined as follows. The experimentally determined overall KIE for the reaction of toluene with OH radical ($^{OH}\varepsilon_{C_7H_8}$) was 5.95‰ ± 0.28‰.[33] Note that all values of KIE stated herein are expressed as epsilon values ($\varepsilon = k_{12}/k_{13} - 1$, where $k_{12}$ is the rate constant for a reactant containing only $^{12}C$ carbons in the molecule and $k_{13}$ is the rate constant for a reactant containing both $^{12}C$ carbons and a $^{13}C$ carbon in the molecule). The $^{OH}\varepsilon_{C_7H_8}$ is the sum of the two channel-specific KIEs:

$$^{OH}\varepsilon_{C_7H_8} = {^{OH}R_{adct}} \times \frac{1}{N_C} \times {^{OH}\varepsilon_{adct}} + {^{OH}R_{abst}} \times \frac{1}{N_C} \times {^{OH}\varepsilon_{abst}} \qquad (2)$$

where $^{OH}R_{abst}$, $^{OH}\varepsilon_{abst}$, $^{OH}R_{adct}$, $^{OH}\varepsilon_{adct}$, and $N_C$ are the branching ratios for H abstraction from a methyl group, the primary carbon KIE for H abstraction from a $^{13}C$ atom in the methyl group, the branching ratio for OH addition to the aromatic carbon ring, the primary carbon KIE for OH addition to a $^{13}C$ atom in the aromatic ring, and the carbon number of reactant molecules, respectively. Note that the KIEs over the $N_C$ times the branching ratios indicate channel-specific primary carbon KIEs, in which the probability of reaction at a $^{13}C$ atom was considered. Methods to estimate the channel-specific KIEs assume no site preference in the distribution of the $^{13}C$ atom within the molecule, which has been validated according to the results of structure-related reactivity analysis as well as systematically changing carbon KIEs (i.e., agreement with the trend of $1/N_C$) for groups of n-alkanes or aromatic hydrocarbons.[33, 34] Using Eq. (2) with the $^{OH}\varepsilon_{abst}$ of 18.7‰ ± 5.2‰[34] and the aforementioned $^{OH}\varepsilon_{C_7H_8}$, the $^{OH}\varepsilon_{adct}$ for OH addition to the toluene ring was calculated to be 44.2‰ ± 5.2‰. This value agrees with the uncertainty range of the estimated $^{OH}\varepsilon_{adct}$ of 40.6‰ ± 1.4‰ from the best fit curve of KIEs versus the number of carbon atoms for a variety of aromatic hydrocarbons.[33] $^{OH}\varepsilon_{adct}$ or $^{OH}\varepsilon_{abst}$ normalized by probabilities of $^{13}C$ distributions in the methyl group and the aromatic ring allows a probability-based KIE to be calculated for each channel of the toluene reaction with OH. By multiplying these values by the branching ratios and the channel-specific $k_{12}$ values, channel-specific $k_{13}$ values were obtained. In addition to the $^{OH}\varepsilon_{abst}$ and the $^{OH}\varepsilon_{adct}$, there was a channel specific KIE for H abstraction from the OH-group of hydroxyaromatic compounds. We assumed this KIE to be negligibly small (~ 0‰) since its impact on $\delta^{13}C$ is a secondary isotope effect. Nevertheless, it cannot be excluded completely that the carbon KIE for this reaction is somewhat larger than expected for a typical secondary isotope effect due to delocalization of the unpaired electron of the phenoxy-group.

However, carbon KIEs for reactions of phenolic compounds with OH radicals have not been measured, and any assumption of a significant KIE would be highly speculative. Parameters and estimated channel specific KIEs used for δ$^{13}$C calculations of photooxidation products are listed in Table 5.

**Table 5. Rate constants, kinetic isotope effects, and branching ratios used for the isotope model**

| branching ratio | channel-specific $^{12}$k (cm$^3$ s$^{-1}$ molecuel$^{-1}$) | channel-specific ε (‰) | channel-specific $^{13}$k (cm$^3$ s$^{-1}$ molecuel$^{-1}$) |
|---|---|---|---|
| 0.88 [a] | 5.2×10$^{-12}$ | 7.5 [b] | 5.2×10$^{-12}$ |
| 0.12 [a] | 7.2×10$^{-13}$ | 0.27 [b] | 7.1×10$^{-13}$ |
| 0.1 | 5×10$^{-17}$ | 0 | 5×10$^{-16}$ |
| 0.9 | 4.9×10$^{-16}$ | 0 | 4.9×10$^{-16}$ |
| 1.0 | 3.6×10$^{-11}$ | 0 | 3.6×10$^{-11}$ |
| 0.9 | 4×10$^{-11}$ | 5.7 [e] | 3.7×10$^{-12}$ |
| 0.1 | 4×10$^{-12}$ | 0 | 4×10$^{-12}$ |
| 0.9 | 5×10$^{-12}$ | 5.7 [e] | 4.5×10$^{-12}$ |
| 0.1 | 5×10$^{-13}$ | 0 | 5×10$^{-13}$ |
| 0.9 | 1.9×10$^{-11}$ | 5.7 [e] | 1.8×10$^{-11}$ |
| 0.1 | 2×10$^{-12}$ | 0 | 2×10$^{-12}$ |

| Reaction channel | overall k (cm³ s⁻¹ molecuel⁻¹) |
|---|---|
| 1a | $5.96 \times 10^{-12}$ [a] |
| 1b | |
| 2a | $(5.4 \pm 0.6) \times 10^{-16}$ [c] |
| 2b | |
| 2c | $4.0 \times 10^{-11}$ [d] |
| 3a | |
| 3b | $4.10 \times 10^{-11}$ [d] |
| 4a | |
| 4b | $5 \times 10^{-12}$ [f] |
| 5a | |
| 5b | $2.05 \times 10^{-11}$ [g] |

[a] The value was adapted from Atkinson.[24] [b] The KIE reported by Anderson et al.[34] was used to estimate the channel specific ε. [c] The second-order k for the reaction of methylcyclohexadienyl radical with O₂ was adopted from Knispel et al. (1990).[30] [d] The value is for the o-cresol + OH reaction (Calvert et al.).[23] [e] The KIE for this channel was estimated according to the generalized carbon KIE for OH adduct reaction to the aromatic ring (40.6‰ ± 1.4‰) divided by $N_c$, reported by Anderson et al.[33] [f] The value was adapted from Bejan et al.[22] [g] The value was adapted from Olariu et al..[32]

Using the parameters listed in Table 5, δ¹³C values of toluene and its photooxidation products were calculated. Note that the rate constants in the table are for the reactions in the gas phase; the rate constants for such reactions at the surface of or inside particles remain unknown. However, these unknown rate constants are expected to be small because SOA experiments for the reaction of β-pinene with ozone showed that ¹³C enrichment in first-generation products in the gas phase was greater than that in SOA,[12] indicating that reactions in the gas phase proceeded faster than those at the surface of or inside particles.

      Predicted δ¹³C values were calculated for methylnitrophenol, the second-generation product via the 3b channel, methylcyclohexadienyl radical, and the sum of all products from the photooxidation of toluene (Figure 4). The calculations were made for two cases: case I, most of the produced methylnitrophenol partitioned into the condensed phase immediately and no further reaction takes place; case II, 90% of produced methylnitrophenol remained in the gas-phase and undergo further reactions, and the remaining methylnitrophenol partitions to the condensed phase, which is under equilibrium with methylnitrophenol in the gas phase. For case II, note that "under equilibrium"

implies that $\delta^{13}C$ values of methylnitrophenol in the gas-phase and the condensed phase are identical. This assumption is based on previous observations that nopinone in both the gas-phase and the condensed phase produced by β-pinene oxidation has very similar $\delta^{13}C$ values [11].

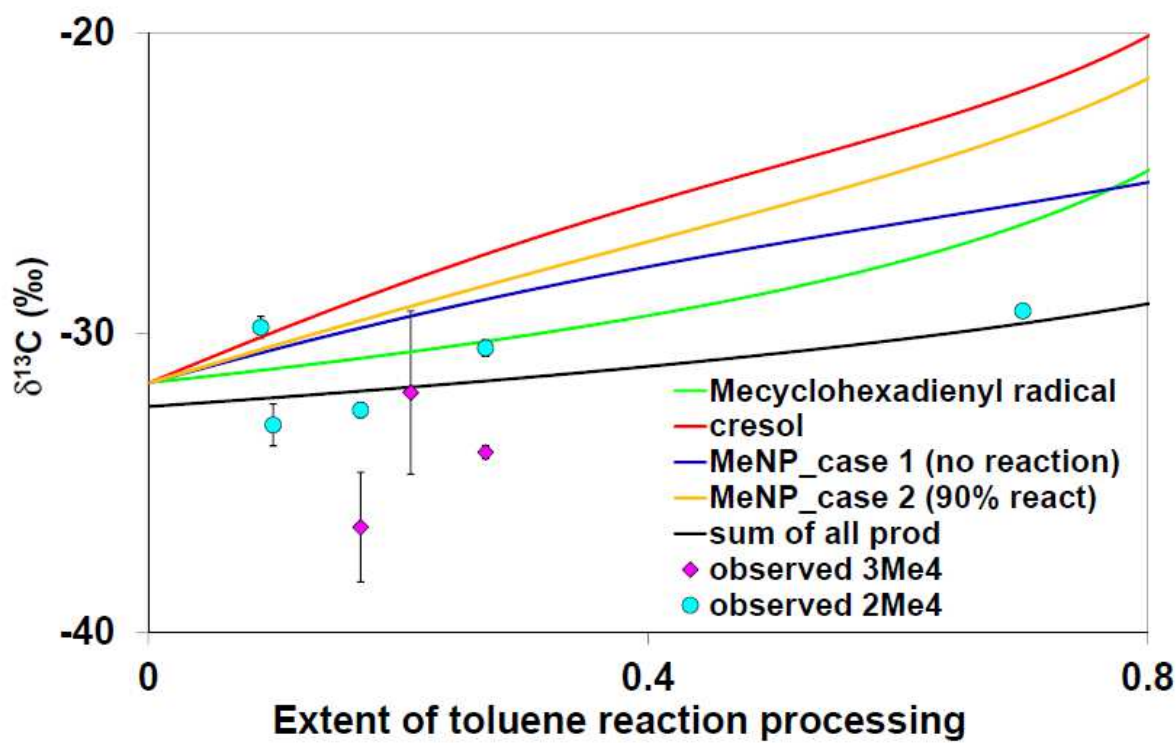

**Figure 4.** Measured and predicted $\delta^{13}C$ values for methylnitrophenol (MeNP). As a comparison, predicted $\delta^{13}C$ values for the methylcyclohexadienyl radical and the sum of all products from toluene photooxidation are also shown

Figure 4 shows that the predicted $\delta^{13}C$ values for methylnitrophenol in case I and case II are similar when toluene turnover is low (up to ~0.3). However, the difference increases at a higher toluene turnover due to the KIE for OH addition to the aromatic ring of methylnitrophenol. The predicted trends for nitrophenols in both cases I and II resulted in two- to three-fold increases in $\delta^{13}C$ compared to observed $\delta^{13}C_{free}$ values for 3-methyl-4- and 2-methyl-4-nitrophenols. Comparison with the observed $\delta^{13}C_{free}$ values for these methylnitrophenols demonstrated that the observations do not agree with the predicted $\delta^{13}C$ values for methylnitrophenols. It remains unclear whether this discrepancy can be attributed to our assumptions on unavailable KIEs (i.e., 0‰ KIEs, as shown in Table 5) or whether it indicated likewise the results of yields that the production is via different reaction channels, such as the 2c channel proposed by Zellner et al.[29] It is possible that the $NO_2$ addition to methylcyclohexadienyl radical forms a $NO_2$ adduct with a high excitation state, which may undergo further reactions with OH, resulting in the formation of methylnitrophenols and methylnitrocatechols. At this time, experimental evidence to support this hypothesis is not available.

Assuming that the majority of methylnitrocatechol partitions into the condensed phase with no loss reaction in the condensed phase, $\delta^{13}C$ values for methylnitrocatechol were calculated for three cases: case A, it is produced via the 4a and the 5b channels; case B, it is produced via the 4a channel only; case C, it is produced via the 5b channel only (Figure 5). Figure 5 also shows predicted $\delta^{13}C$ values for the methylcyclohexadienyl radical, methylcatechol, and methylnitrophenol as comparison. Note that in these calculations, the majority (90%) of the methylnitrocatechol precursor, methylnitrophenol, was assumed to react in the gas-phase and produce particulate methylnitrocatechol. As discussed earlier, the observed $\delta^{13}C_{free}$ values for methylnitrocatechol isomer 1 agree with the predicted $\delta^{13}C$ values for the sum of all products within $2\sigma$, but disagree with the predicted $\delta^{13}C$ for methylnitrocatechol in all three cases. The observed $\delta^{13}C_{free}$ values for methylnitrocatechol isomer 2 and 3 show larger fluctuations, but are overall slightly lower than isomer 1 and closer to the predicted $\delta^{13}C$ for methylnitrocatechol. Unfortunately, the number of observations in this limited range of toluene turnover is insufficient to conclude whether they follow the predicted $\delta^{13}C$ trends for methylnitrocatechol.

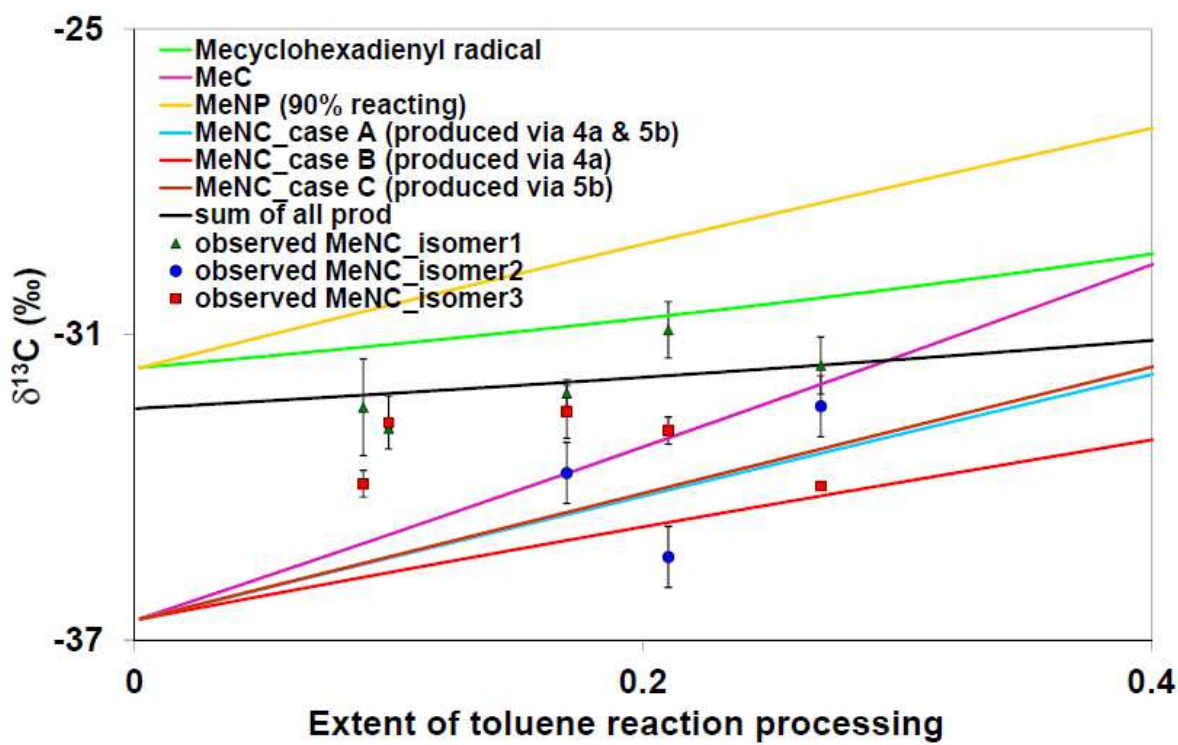

**Figure 5.** Measured and predicted $\delta^{13}C$ values for methylnitrocatechol (MeNC) produced through three pathways (4a channel, 5b channel, or a combination of 4a and 5b channels). As a comparison, predicted $\delta^{13}C$ values for methylcyclohexadienyl radicals, as well as the sum of all products from toluene photooxidation, methylcatechol (MeC), and methylnitrophenol (MeNP) are shown

Figure 6 shows the mechanism for production of nitrophenol proposed by Atkinson.[14] For the production of 4-nitrophenol, the reaction mechanism involves a reaction at the methyl group of toluene; therefore, production of 4-nitrophenol must proceed via the 1b channel in Figure 2. Subsequent reactions are shown in detail in Figure 6. According to the mechanism, significant carbon isotope fractionations are expected at the initial step (i.e., H abstraction from the methyl group) when benzaldehyde reacts with OH radical (i.e., abstraction of a hydrogen atom from the aldehyde group) because both reactions are cleavages of primary C–H bonds (carbon KIEs). These fractionations result in $^{13}C$ depletion in the reaction products. However, this carbon isotope fractionation occurs in the methyl or aldehyde carbon atom, which is lost as the intermediate benzaldehyde is further oxidized (i.e., decarboxylated). As a consequence of decarboxylation, the $\delta^{13}C$ for the formed 4-nitrophenol is almost identical to that for the aromatic ring of toluene, with a $\delta^{13}C$ value of ~ –27‰ (Figure 7). The mechanism proposed for 4-nitrophenol is consistent with the observed isotope ratios.

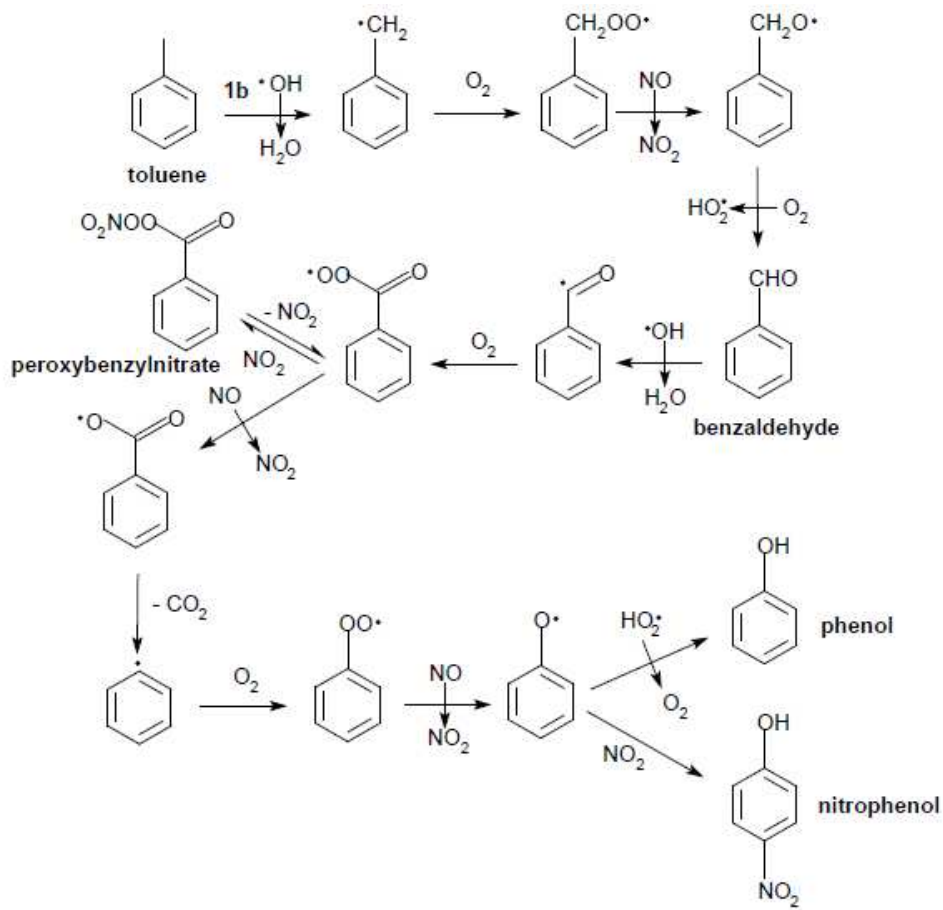

**Figure 6.** Postulated reaction mechanisms for production of nitrophenol via 1b channel.[14]

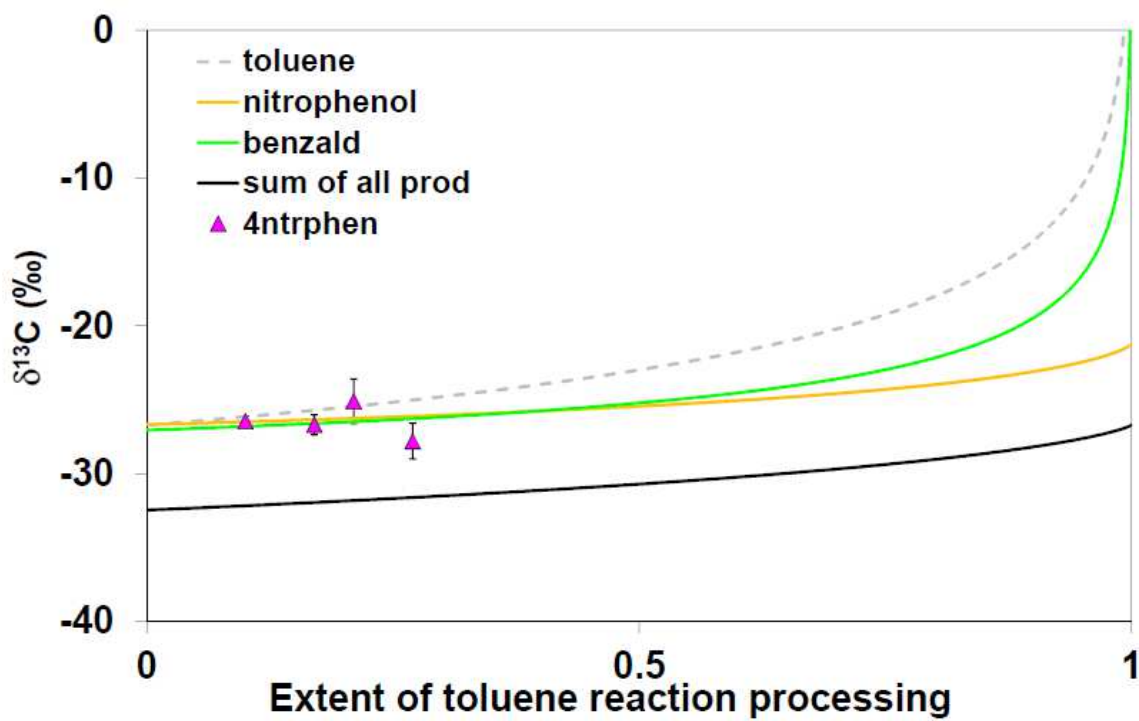

**Figure 7.** Measured $\delta^{13}C$ value for 4-nitrophenol and predicted $\delta^{13}C$ values for toluene, benzaldehyde, nitrophenol, and the sum of all products.

## 4. SUMMARY AND CONCLUSION

GC-MS analysis for SOA extracts formed by photooxidation of toluene identified 8 phenolic compounds. The most abundant product was 2-methyl-4-nitrophenol. Compound-specific stable carbon isotope ratios ($\delta^{13}C$) for the identified phenolic compounds demonstrated that the observed $\delta^{13}C$ values for 4-nitrophenol were the similar to the initial $\delta^{13}C$ of toluene, while the $\delta^{13}C$ values of the other phenolic compounds were depleted by 5‰–6‰ compared to the initial $\delta^{13}C$ value of toluene. For 4-nitrophenol, the $\delta^{13}C$ value dependence on precursor conversion was consistent with the formation mechanism proposed in previous reports. However, the observed $\delta^{13}C$ values for the other phenolic compounds were significantly lower than $\delta^{13}C$ values estimated for these products according to the postulated mechanisms (i.e., production via the 3a and the 3b channels). Although the observed $\delta^{13}C$ values varied, they were similar to the predicted $\delta^{13}C$ for the sum of all products using the mass balance. This suggests that these particulate phenolic products were produced via different reaction channels. To constrain production channels, more experimental evidence, such as KIE studies for intermediates, are required. Specifically, the lack of measured KIEs for reactions of phenols with OH-radicals creates uncertainty in predicted $\delta^{13}C$ values for phenolic compounds by the oxidation of toluene in air. Nevertheless, our results demonstrate the usefulness of compound-specific stable carbon isotope measurements to increase our understanding of chemical reaction mechanisms.

**ACKNOWLEDGMENTS**

We thank Darrell Ernst, Alina Chivulescu, and Wendy Zhang at Environment Canada for technical support. We also acknowledge the Natural Sciences and Engineering Research Council of Canada, the Canadian Foundation for Climate and Atmospheric Sciences, and Environment Canada for financial support. The project was also partially supported by the ESPEC Foundation for Global Environment Research and Technologies (Charitable Trust).

**Supporting Information**

Retention time index and reference $\delta^{13}C$ values for standard compounds (Table S-1); a GCC-IRMS chromatogram for a blank extract (Figure S-1) are also presented.